\documentclass[twocolumn, 
  aps, pra,
  amsmath,amssymb,
  longbibliography,
  ]{revtex4-1}
\usepackage{graphicx,color}
\usepackage{amsmath}
\usepackage{natbib}
\usepackage{epsfig}
\begin{document}

\title{\color{blue} Thermal conductivity of strongly coupled Yukawa fluids}

\author{Sergey A. Khrapak}
\email{Sergey.Khrapak@gmx.de}
\affiliation{Joint Institute for High Temperatures, Russian Academy of Sciences, 125412 Moscow, Russia 
}

\begin{abstract}
A vibrational model of heat conduction in liquids with soft pairwise interactions is applied to estimate the thermal conductivity coefficient of strongly coupled Yukawa fluids. A reasonable agreement with the available data from numerical simulations is observed. The results can be useful in the context of strongly coupled plasma and complex (dusty) plasma fluids, when Yukawa (or screened Coulomb) interaction potential is applicable.
\end{abstract}

\date{\today}

\maketitle


A simple vibrational model of heat transfer in simple liquids with soft pairwise interatomic interactions has been put forward recently~\cite{KhrapakPRE01_2021}. In this model the thermal conductivity coefficient is expressed via an effective frequency of atomic vibrations around a local temporary equilibrium position, mean interatomic separation and specific heat capacity. The vibrational frequency is determined by averaging over the liquid collective mode excitation spectrum. The model has been applied to quantify heat transfer in a dense Lennard-Jones liquid and a strongly coupled one-component plasma (OCP). A remarkable agreement with the available numerical results has been documented~\cite{KhrapakPRE01_2021}. The purpose of this Brief Communication is to apply the model to strongly coupled Yukawa fluids and to examine its accuracy in this case.   

In the vibrational model of thermal conduction, a liquid is approximated by a layered structure with layers perpendicular to the temperature gradient and separated by a distance $\Delta = n^{-1/3}$, where $n$ is the particle (or atomic) density. The average interparticle separation within such quasi-layers is also $\Delta$.  The particle positions in the liquid's quasi-layers are not regular and are not fixed. The particles can migrate within a given layer as well as between different layers. However, the time scale of these migrations is relatively long in dense liquids, much longer than the characteristic time of latticle-like vibrations of a particle around its temporary equilibrium position~\cite{FrenkelBook}. This resembles the applicability condition of the quasi-localized charge approximation (QLCA), which is very successively used to describe collective modes in strongly coupled plasma-related systems~\cite{GoldenPoP2000}.

The average difference in energy between the atoms of adjacent layers is $\Delta (dU/dx)=\Delta c_{\rm v}(dT/dx)$, where $U$ is the internal energy and $c_{\rm v}=dU/dT$ is the specific heat at constant volume. Although the specific heat at constant pressure, $c_{\rm p}$, would be apparently more appropriate for such an estimate, the difference between $c_{\rm p}$ and $c_{\rm v}$ is vanishingly small in strongly coupled Yukawa fluids. At the same time $c_{\rm v}$ is easier to evaluate (see below) so it is appropriate to use $c_{\rm v}$ for practical purposes. The energy exchange between neighbouring quasi-layers occurs at a  characteristic frequency $\nu$, which is set equal to the average vibrational frequency of a particle, $\nu=\langle \omega \rangle/2\pi$. The energy flux between two layers per unit area is ${\mathcal E}=-\nu\Delta c_{\rm v}(dT/dx)/\Delta^{2}$. This should be compared with the Fourier law, ${\mathcal E}=-\lambda(dT/dx)$, resulting in  
\begin{equation}\label{lambdaR}
\lambda=c_{\rm v}\frac{\langle \omega \rangle}{2\pi \Delta}.
\end{equation}
This is a simple general expression without free parameters, which we are going to examine in detail using strongly coupled Yukawa fluids.  

Since the actual frequency distribution can be quite complex in liquids, and can vary from one type of liquid to another, some simplifying assumptions have to be employed in evaluating $\langle \omega \rangle$. Before we do that for Yukawa fluids, let us point out two special cases of Eq.~(\ref{lambdaR}). In the simplest Einstein approximation all atoms vibrate with the same (Einstein) frequency $\Omega_{\rm E}$ so that $\langle \omega \rangle = \Omega_{\rm E}$. We obtain the expression proposed by Horrocks and McLaughlin (to within some difference in numerical coefficients)~\cite{Horrocks1960}
\begin{equation}\label{Horrocks}
\lambda = c_{\rm v} \frac{\Omega_{\rm E}}{2\pi \Delta}.
\end{equation}

As a more involved approximation, assume that a dense liquid supports one longitudinal and two transverse modes, which exhibit acoustic dispersion relations in the long-wavelength limit. The Debye-like averaging procedure yields~\cite{KhrapakPRE01_2021}   
\begin{equation}\label{Debye}
\lambda \simeq \frac{1}{4}\left(\frac{3}{4\pi}\right)^{1/3} c_{\rm v} \frac{c_l+2c_t}{\Delta^2}, 
\end{equation}
where $c_l$ and $c_t$ are the longitudinal and transverse sound velocities, respectively. This is similar (again to within numerical coefficients) to a formula proposed by Cahill and Pohl to describe thermal conductivity coefficient of amorphous inorganic solids~\cite{Cahill1989,Cahill1992}. 

The Yukawa systems considered here represent a collection of point-like charged particles immersed into a neutralizing medium (usually conventional electron-ion plasma), which provides screening. The pairwise Yukawa repulsive interaction  potential (also known as screened Coulomb or Debye-H\"uckel potential) is
\begin{equation}\label{Yukawa}
\phi(r)=(Q^2/r)\exp(-\kappa r/a),
\end{equation}
where $Q$ is the particle charge and $\kappa$ is the dimensionless screening parameter, which is the ratio of the Wigner-Seitz radius $a=(4\pi n/3)^{-1/3}$ to the plasma screening length. Yukawa potential is widely used as a reasonable first approximation for actual  interactions in three-dimensional isotropic complex plasmas and colloidal suspensions~\cite{TsytovichUFN1997,FortovUFN,FortovPR,
KhrapakPRL2008,KhrapakCPP2009,ChaudhuriSM2011,IvlevBook,
LampePoP2015}. 

The dynamics and thermodynamics of Yukawa systems are conventionally characterized by the screening parameter $\kappa$ and the coupling parameter $\Gamma=Q^2/aT$, where 
$T$ is the system temperature (in energy units).
The coupling parameter characterizes the ratio between the potential energy of interparticle interaction and the kinetic energy. In strongly coupled Yukawa fluids the condition $\Gamma\gg 1$ should be satisfied. For even higher coupling Yukawa fluids crystallize, forming either the body-centered-cubic (bcc) or face-centered-cubic (fcc) lattices. Detailed phase diagrams of Yukawa systems are available in the literature~\cite{RobbinsJCP1988,HamaguchiJCP1996,HamaguchiPRE1997,
VaulinaJETP2000,VaulinaPRE2002}. 
The screening parameter $\kappa$ determines the softness of the interparticle repulsion. It varies from the extremely soft and long-ranged Coulomb potential at $\kappa\rightarrow 0$ (OCP limit) to the hard-sphere-like interaction limit at $\kappa\rightarrow \infty$. In the context of complex plasmas and colloidal suspensions the relatively ``soft'' regime, $\kappa\sim {\mathcal O}(1)$, is of particular interest. Most of previous investigations have focused on this regime and we follow this tradition. 

Transport properties (in particular, self-diffusion, shear viscosity, and thermal conductivity coefficients) of Yukawa fluids have been relatively well investigated. Thermal conduction in three-dimensional Yukawa fluids has been studied and discussed in Refs.~\cite{SalinPRL2002,SalinPoP2003,FaussurierPRE2003,
DonkoPRE2004,VaulinaPRE2010,ShahzadPoP2012,
OttPRE2015,ScheinerPRE2019,KahlertPPR2020}. For a detailed comparison we chose the data from non-equilibrium molecular dynamics simulations reported by Donko and Hartmann~\cite{DonkoPRE2004}. This is the most complete set of data in terms of covered $\Gamma$ and $\kappa$ parameters from those available presently (note that there exists some deviations between the results from different investigations as shown for instance in Fig.~5 of Ref.~\cite{KahlertPPR2020}). Throughout this paper we employ Rosenfeld's normalization~\cite{RosenfeldJPCM1999} for the thermal conductivity coefficient $\lambda_{\rm R}=\lambda n^{-2/3}/v_{\rm T}$ (where $v_{\rm T}=\sqrt{T/m}$ is the thermal velocity), which is different from that used originally, $\lambda'=\lambda/(n\omega_{p}a^2)$, where $\omega_{\rm p}=\sqrt{4\pi Q^2 n/m}$ is the plasma frequency scale. Note that since we express temperatures in energy units, the Boltzmann constant does not show up in the expressions (effectively $k_{\rm B}=1$). 

\begin{figure}
\includegraphics[width=8.cm]{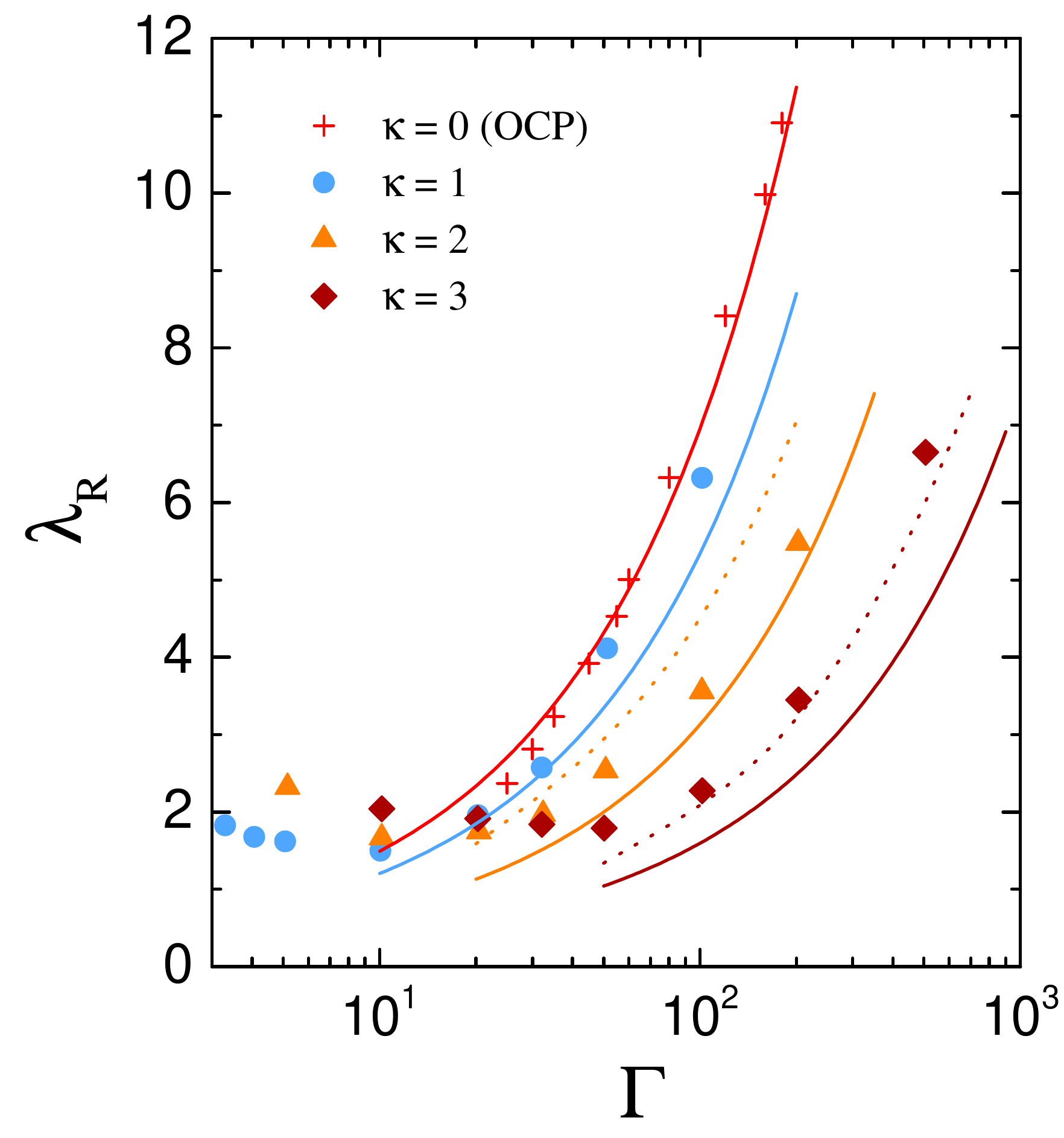}
\caption{(Color online) Reduced thermal conductivity coefficient $\lambda_{\rm R}$ versus the coupling parameter $\Gamma$. Citcles, triangles, and rhombs correspond to molecular dynamics results from Ref.~\cite{DonkoPRE2004} for $\kappa=1$, $\kappa=2$, and $\kappa=3$, respectively. Crosses are the OCP numerical results from Ref.~\cite{ScheinerPRE2019}. The solid curves are the theoretical calculations using Eq.~(\ref{lambdaR}). The dotted curves (shown for $\kappa=2$ and $\kappa=3$) are plotted using Eq.~(\ref{Debye}). The results of calculation using Eq.~(\ref{Horrocks}) are not shown for clarity, but are discussed in the text instead.}
\label{Fig1}
\end{figure}   

The numerical data for the dependence of $\lambda_{\rm R}$ on $\Gamma$ are plotted in Fig.~\ref{Fig1}.  The solid symbols correspond to the data from a non-equilibrium simulation in Ref.~\cite{DonkoPRE2004} for strongly coupled Yukawa fluids with $\kappa=1.0$, $2.0$, and $3.0$. The crosses correspond to the OCP results from Ref.~\cite{ScheinerPRE2019}, added for completeness. The minima in $\lambda_{\rm R}$ separate the regions of gaseous-like and fluid-like regimes of thermal conductivity. We will consider only the fluid-like (i.e. strongly coupled) regime to the right from the minima.   
  
\begin{figure*}
\includegraphics[width=15.cm]{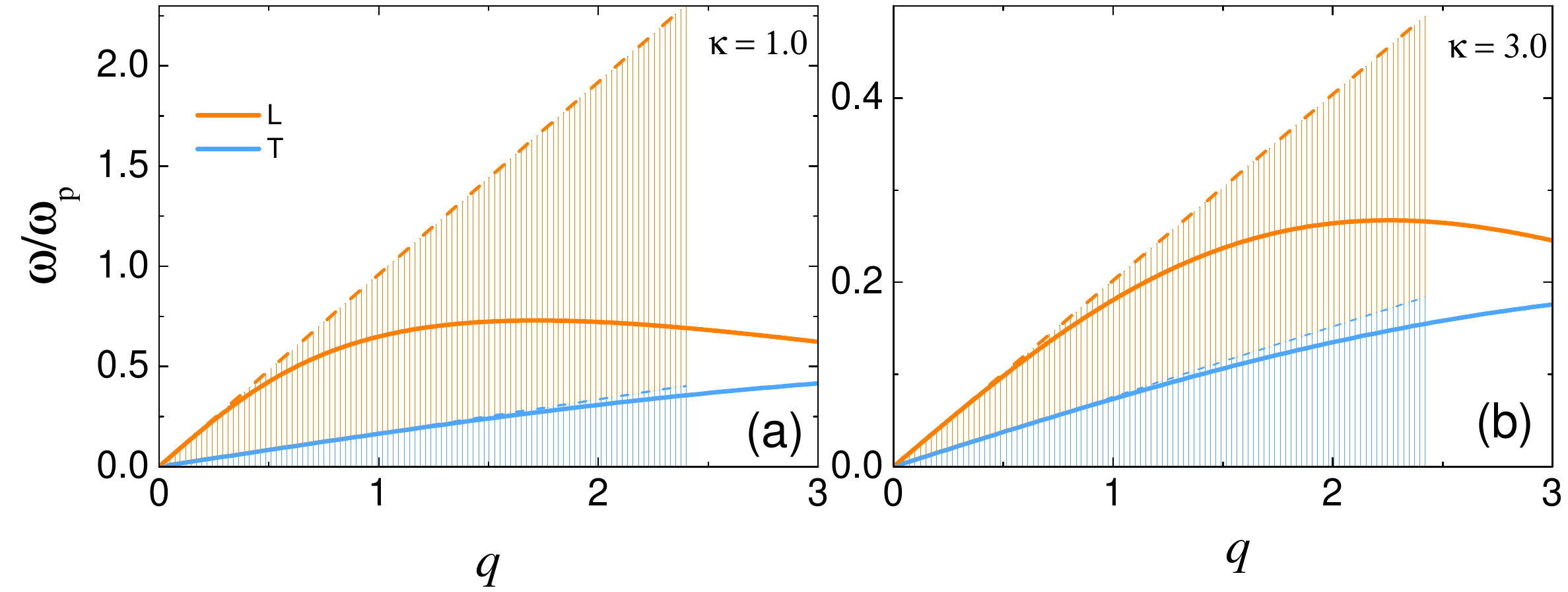}
\caption{(Color online) Dispersion relations of strongly coupled Yukawa fluids with $\kappa=1.0$ (a) and $\kappa=3.0$ (b). The wave frequency $\omega$ is expressed in units of the plasma frequency $\omega_{\rm p}$, the reduced wave-number is $q=ka$. The solid curves correspond to the simplified QLCA dispersion relations of Eq.~(\ref{L1}) for the longitudinal mode and Eq.~(\ref{T1}) for the transverse mode. Dashes lines are the corresponding acoustic asymptotes. They terminate at $q_{\rm max}=(9\pi/2)^{1/3}$, the maximum cutoff wavelength involved in the averaging procedure. Note different scales of the vertical axis in (a) and (b).}
\label{Fig2}
\end{figure*} 

To get the average vibrational frequency, we perform the conventional averaging over collective modes that are supported in a strongly coupled (dense) fluid. Taking into account one longitudinal and two transverse modes the average frequency becomes~\cite{KhrapakPRE01_2021}
\begin{equation}\label{integral}
\langle \omega \rangle = \frac{2}{9\pi}\int_0^{\rm q_{\rm max}}q^2dq\left[\omega_l(q)+2\omega_t(q)\right],
\end{equation} 
where $q=ka$ is the reduced wave number and $\omega_{l,t}(q)$ are the dispersion relations of the longitudinal and transverse modes. The cutoff wave number $q_{\rm max}$ appears due to the normalization condition, which should ensure that each collective mode contains $n$ oscillations. Mathematically this implies that 
\begin{displaymath}
\frac{4\pi}{3}\left(\frac{k_{\rm max}}{2\pi}\right)^3=n; \quad \Rightarrow \quad q_{\rm max}=\left(\frac{9\pi}{2}\right)^{1/3}\simeq 2.418.
\end{displaymath}
This cutoff also ensures that averaging of a quantity that does not depend on $q$ will not change its actual value.
The longitudinal and transverse collective mode frequencies can be well approximated within the QLCA approach, which relates the dispersion relations to the pairwise interaction potential and structural characteristics, expressed in the form of the radial distribution function (RDF) $g(r)$~\cite{GoldenPoP2000}.
Additionally, we make use of a simple excluded cavity model for the fluid RDF~\cite{KhrapakPoP2016}.       
This results in simple and elegant expressions for the  dispersion relations of the longitudinal and transverse modes~\cite{KhrapakPoP2016}:
\begin{equation}\label{L1}
\begin{aligned}
\omega_{l}^2=\omega_{\rm p}^2e^{-R\kappa}\left[\left(1+R\kappa\right)\left(\frac{1}{3}-\frac{2\cos Rq}{R^2q^2}+\frac{2\sin Rq}{R^3q^3} \right) \right. \\ \left. -\frac{\kappa^2}{\kappa^2+q^2}\left(\cos Rq+\frac{\kappa}{q}\sin Rq \right)\right],
\end{aligned}
\end{equation}  
and
\begin{equation}\label{T1}
\omega_t^2=\omega_{\rm p}^2e^{-R\kappa}\left(1+R\kappa\right)\left(\frac{1}{3}+\frac{\cos Rq}{R^2q^2}-\frac{\sin Rq}{R^3q^3} \right).
\end{equation}
Here the excluded cavity [the region where $g(r)=0$] radius $R$ is expressed in units of $a$ and is related to the screening parameter via~\cite{KhrapakAIPAdv2017,KhrapakIEEE2018}
\begin{equation}\label{R}
R(\kappa)\simeq 1+\frac{1}{\kappa}\ln \left[\frac{3 \cosh (\kappa)}{\kappa^2}-\frac{3 \sinh
(\kappa)}{\kappa^3}\right].
\end{equation}
These fully analytical dispersion relations of strongly coupled Yukawa fluids demonstrate a very high accuracy in the long-wavelength regime~\cite{KhrapakPoP2016,KhrapakAIPAdv2017,KhrapakIEEE2018}. 
The long-wavelength portion of these dispersion relations for $\kappa=1$ and $\kappa=3$ are plotted in Fig.~\ref{Fig2}. Note that although the existence of a so-called ``$q$-gap'' (zero frequency non-propagating branch at long wavelengths~\cite{MurilloPRL2000,GoreePRE2012,TrachenkoRPP2015,KhrapakJCP2019,
KryuchkovSciRep2019}) in the transverse mode is not reproduced, this does not affect our main results significantly, because the contribution from the long wavelengths (low frequencies) to the thermal conductivity is relatively small.

We can now compare simulations results with theoretical approximations of Eqs. (\ref{lambdaR}), (\ref{Horrocks}), and (\ref{Debye}). 

The integral in Eq.~(\ref{integral}) has been evaluated numerically for different screening parameters and the resulting values for $\langle \omega\rangle/\omega_{\rm p} $ are presented in Table~\ref{Tab1}. In the same table the values of $\Omega_{\rm E}/\omega_{\rm p}$ needed to evaluate Eq.~(\ref{Horrocks}) are also given. Here we use the fact that the ratio $\Omega_{\rm E}/\omega_{\rm p}$ is virtually independent of $\Gamma$ in the strongly coupled regime (be it fluid or crystal)~\cite{KhrapakPoP2018} and use the fcc data tabulated by Ohta and Hamaguchi~\cite{OhtaPoP2000}. Note that the dependence on the coupling parameter is absorbed in the plasma frequency. Also quoted in the Table~\ref{Tab1} are the values $\sqrt{\langle\omega^2\rangle/\omega_{\rm p}^2}$ evaluated inserting $\omega_l^2(q)+2\omega_t^2(q)$ under the integral in Eq.~(\ref{integral}). Very close agreement with the values $\Omega_{\rm E}/\omega_{\rm p}$ demonstrates that all the approximations made to deal with dispersion relations of strongly coupled Yukawa fluids are appropriate.          

The longitudinal and transverse sound velocities can be evaluated from 
\begin{equation}
c_{l,t}=\lim_{q\rightarrow 0}\frac{\omega_{l,t}(q)a}{q}.
\end{equation}
Simple expressions are available~\cite{KhrapakIEEE2018}, and these were used to evaluate $c_l$ and $c_t$ to be substituted in Eq.~(\ref{Debye}).

\begin{table}
\caption{\label{Tab1} The average of vibrational frequencies $\langle\omega \rangle$ and $\sqrt{\langle\omega^2 \rangle}$ of strongly coupled Yukawa fluids as well as the Einstein frequency $\Omega_{\rm E}$ of the fcc Yukawa solid for screening parameters up to $\kappa=5$. All frequencies are reduced using the plasma particle frequency.}
\begin{ruledtabular}
\begin{tabular}{lccccc}
$\kappa$ & 1.0 &  2.0 & 3.0 & 4.0 &  5.0  \\ \hline
$\langle\omega \rangle/\omega_{\rm p}$ & 0.420 & 0.279 & 0.164 & 0.088 & 0.044    \\
$\sqrt{\langle\omega^2 \rangle}/\omega_{\rm p}$ & 0.469 & 0.305 & 0.176 & 0.094 & 0.0469    \\
$\Omega_{\rm E}/\omega_{\rm p}$& 0.472 & 0.307 & 0.176 & -- & 0.0469  \\
\end{tabular}
\end{ruledtabular}
\end{table}  

The last remaining step is to evaluate the specific heat $c_{\rm v}$. Thermodynamics of Yukawa fluids has been relatively well investigated~\cite{HamaguchiPRE1997,FaussurierPRE2003,ToliasPRE2014,
KhrapakPRE02_2015,KhrapakPPCF2015,
KhrapakJCP2015,ToliasPoP2015,ToliasPoP2019}. Various practical expressions for thermodynamic functions  have been proposed in the literature. Among these, the freezing temperature scaling of the thermal component of the internal energy proposed by Rosenfeld and Tarazona~\cite{RosenfeldMolPhys1998,RosenfeldPRE2000} combines  relative good accuracy with simplicity and relatively wide applicability~\cite{IngebrigtsenJCP2013}. In this apprximation the thermal correction to the internal energy of Yukawa fluids (the latter is dominated by the static contribution for sufficiently soft potentials, like Yukawa considered here) scales as $\propto T^{3/5}$ and hence the reduced correction is $\propto T^{-2/5}\propto \Gamma^{2/5}$. In the range of $\kappa$ considered the reduced thermal correction per particle can be well approximated as $u_{\rm th}\simeq 3.1 (\Gamma/\Gamma_{\rm fr})^{2/5}$~\cite{KhrapakJCP2015}, where $\Gamma_{\rm fr}$ is the value of the coupling parameter at freezing, which is very close to that at melting, $\Gamma_{\rm m}$ (the coexistence region is very narrow, so that usually there is no need to distinguish between $\Gamma_{\rm fr}$ and $\Gamma_{\rm m}$ for Yukawa systems). The corresponding specific heat at constant volume is therefore
\begin{equation}\label{cv}
c_{\rm v}\simeq 1.5+1.86\left(\frac{\Gamma}{\Gamma_{\rm fr}}\right)^{2/5}.
\end{equation} 
The values of $\Gamma_{\rm fr}$ for various $\kappa$ have been obtained and tabulated in Ref.~\cite{HamaguchiPRE1997}  (relatively accurate fits are also available~\cite{VaulinaJETP2000,VaulinaPRE2002}.) We now have all the necessary information to evaluate each of the equations (\ref{lambdaR}), (\ref{Horrocks}), and (\ref{Debye}). 

A comparison between the vibrational model of heat conduction and the results from numerical experiments is shown in Fig.~\ref{Fig1}. The solid curves are plotted using the general expression (\ref{lambdaR}), supplemented by the average vibrational frequencies from Table~\ref{Tab1} and the specific heat from Eq.~(\ref{cv}). There is a reasonable agreement with numerical results in particular in the weak screening regime with smaller $\kappa$. In the OCP case we have performed averaging with the help of Eqs.~(\ref{L1}) and (\ref{T1}) with $\kappa=0$~\cite{KhrapakPoP2016}, but used the accurate OCP equation of state to evaluate $c_{\rm v}$ (for further details see Ref.~\cite{KhrapakPRE01_2021}). The agreement is excellent. Note that in the near-OCP regime with $\kappa\lesssim 1$ the thermal conductivity coefficient exhibits a very weak dependence on $\kappa$ (thus OCP result can serve as a relevant approximation in this regime). 

The dotted curves have been calculated using Eq.~(\ref{Debye}). This approximation leads to some overprediction of the thermal conductivity coefficient. The origin of this overprediction is immediately clear from Fig.~\ref{Fig2}. The acoustic asymptotes are overestimating the actual frequencies, in particular for the longitudinal collective mode. The situation gets particularly worse at weak screening. Here the longitudinal sound velocity diverges as $\propto \kappa^{-1}$~\cite{KhrapakPoP2019} (in the OCP limit the longitudinal mode exhibits the plasmon dispersion so that formally $c_l\rightarrow \infty$). This produces grossly inaccurate results at weak screening and we do not show the results of calculation using Eq.~(\ref{Debye}) for $\kappa=1$. For $\kappa=2$ this approximation somewhat overestimate the numerical data and for $\kappa=3$ it is located rather close to the numerical results. Thus, the averaging over a Debye-like acoustic spectrum is inappropriate in the regime of soft long-ranged potential, but becomes more appropriate when the potential becomes steeper (see Fig.~\ref{Fig2} for illustration). It is not surprising, therefore, that this approximation works rather well in the case of Lennard-Jones liquids with a rlatively steep interaction potential~\cite{KhrapakPRE01_2021}.        

The calculation using Eq.~(\ref{Horrocks}) is not shown in Fig.~\ref{Fig2} for clarity, but actually this is a very good approximation for the regime considered. From the Table~\ref{Tab1} we see that the values $\Omega_{\rm E}$ are somewhat higher that those of $\langle \omega\rangle$. This implies that for finite $\kappa$ this approximation would be closer to the numerical results than the general expression (\ref{lambdaR}). In the OCP limit it only slightly overestimates the numerical results, as has been demonstrated previously~\cite{KhrapakPRE01_2021}.

The main conclusion can be formulated as follows. The vibrational model of heat transfer works reasonably well for strongly coupled Yukawa fluids. This model relates the coefficient of thermal conductivity to the specific heat, mean inter-atomic separation and an average vibrational frequency of atoms around their temporary equilibrium positions. The general formula, which uses averaging over collective modes of dense liquids, and its special case, which assumes that all atoms vibrate with the same Einstein frequency, are reasonably accurate in the entire regime investigated. The Debye-like averaging, which uses long-wavelength acoustic asymptotes, is not appropriate near the OCP limit, but becomes appropriate as the steepness of the interaction potential (regulated by the screening parameter $\kappa$) increases. 

The model can be straightforwardly generalized to heat transfer in two-dimensional (2D) Yukawa layers and this has been recently discussed in Ref.~\cite{KhrapakPoP2021}. In the context of  2D complex (dusty) plasma fluids and solids in ground based laboratories, the regime with $\kappa\lesssim 1$ is of particular practical interest. Based on the results of the present analysis it is expected that estimations for the (2D) OCP limit should be relevant in this regime and the Einstein model should be appropriate as well.

Data sharing is not applicable to this article as no new data were created. 



\bibliography{TC_Ref}

\end{document}